\documentclass[usenatbib,usegraphicx]{mn2e}
\onecolumn
\usepackage{subfigure}
\usepackage{longtable}
\usepackage{graphicx}
\usepackage{graphics}
\usepackage{keyval}
\usepackage{trig}

\def\beq{\begin{equation}}
\def\eeq{\end{equation}}
\def\bey{\begin{eqnarray}}
\def\eey{\end{eqnarray}}

\def\msun{M_\odot}

\def\kms{\, {\rm km \, s}^{-1} }

\def\grad{{\bf \nabla}}

\def\a0{$a_0$}

\title{Can MOND take a bullet? Analytical comparisons of three versions of MOND beyond spherical symmetry}
\author[G. W. Angus, B. Famaey and H. S. Zhao]{G. W. Angus$^{1}$\thanks{email:
gwa2@st-andrews.ac.uk}, B. Famaey$^{2}$\thanks{FNRS Scientific
    Collaborator; email:
bfamaey@ulb.ac.be} and H. S. Zhao$^{1}\thanks{email:
hz4@st-andrews.ac.uk}$ \\
$^{1}$SUPA, School of Physics and Astronomy, University of St. Andrews, Scotland KY16 9SS\\
$^{2}$Institut d'Astronomie et d'Astrophysique, Universit\'e Libre de Bruxelles,
CP 226, Boulevard du Triomphe, B-1050 Bruxelles, Belgium\\}

\begin{document}

\date{Accepted ... Received ... ; in original form ...}

\pagerange{\pageref{firstpage}--\pageref{lastpage}} \pubyear{2006}

\maketitle

\label{firstpage}

\begin{abstract}
A proper test of Modified Newtonian Dynamics (MOND) in systems of non-trivial
geometries depends on modelling subtle differences in several versions of its postulated theories.  
This is especially important for lensing and dynamics of barely virialised 
galaxy clusters with typical gravity of scale $\sim a_0 \sim 1\AA{\rm s}^{-2}$.  The original MOND formula, 
the classical single field modification of the Poisson equation, and
the multi-field general relativistic theory of Bekenstein (TeVeS) all lead to different predictions as we stray from spherical symmetry.  In this paper, we
study a class of analytical MONDian models for a system with a 
semi-Hernquist baryonic profile.  After presenting 
the analytical distribution function of the baryons in spherical limits,
we develop orbits and gravitational lensing of the models 
in non-spherical geometries.  
In particular, we can generate a multi-centred baryonic system
with a weak lensing signal resembling that of the merging galaxy cluster
1E 0657-56 with a bullet-like light distribution.    
We finally present analytical scale-free highly non-spherical models to  
show the subtle differences between the single field
classical MOND theory and the multi-field TeVeS theory.    
\end{abstract}

\begin{keywords}
gravitation - dark matter - galaxy: kinematics and dynamics, structure
\end{keywords}

\section{Introduction}
\protect\label{sec:intr}
While luminous models for the central parts of galaxies do not usually
require a dark matter (DM) component, massive halos of DM
must be taken into account if one wants to construct  realistic
potential-density pairs for an entire galaxy, in order to reproduce the
observed nearly-flat galactic rotation curves.

Curiously, there is a considerable body of evidence that the galactic mass
profiles of baryonic and dark matter are not uncorrelated (McGaugh 2005).
The correlation between the Newtonian gravity of the baryons ${\bf g}_N$
and the overall gravity ${\bf g}$ (baryons plus DM) can be loosely
parameterized by Milgrom's (1983) empirical relation
\begin{equation}
\protect\label{eqn:mueq}
        {\mu}(g/a_0) \, {\bf g} = {\bf g}_{{\rm N}},
\end{equation}
where the interpolating function $\mu(x)$ is a function which runs smoothly from
$\mu(x)=x$ at $x\ll 1$ to $\mu(x)=1$ at $x\gg 1$ with a dividing gravity scale 
$a_0 \sim 1 {\rm \AA} {\rm s}^{-2}$
at the transition.  It is a very serious challenge for cold dark matter simulations to
reproduce this empirical relation. On the other hand, this relation can be interpreted as a modification of the Newton-Einstein gravitational
law in the ultra-weak field regime, with no actual need for dark matter (at the galaxy scale). This
provocative idea was taken as the basis for
the MOND theory: however, using Eq.(1) alone to transform 
${\bf g}_{{\rm N}}$ into ${\bf g}$ does not provide a respectable gravitational
force preserving energy and angular momentum. 

Bekenstein \& Milgrom (1984) then suggested modifying the Poisson Equation
in order to produce a non-relativistic MOND potential. This
proposal was recently refined by Bekenstein (2004) who
presented a Lorentz-covariant theory of gravity, dubbed TeVeS, yielding MONDian behaviour in the appropriate limit. However, there is a subtle
difference between the non-relativistic formulation of MOND and the
relativistic TeVeS, namely the number of fields involved: in MOND, the potential is modified directly in order to trigger the MOND
phenomenology, while in TeVeS, a scalar field is added to the traditional
Newtonian potential. These two descriptions
are equivalent only in highly symmetric systems (spherical or cylindrical symmetry).
For disk galaxies, Eq.(\ref{eqn:mueq}) is a good approximation of the MOND and TeVeS
gravitational theories since the additional curl
field (see \S \ref{sec:montev}) is small when solving the modified Poisson equation for the potential
or the scalar field (Brada \& Milgrom 1995). For this
reason, Eq.(\ref{eqn:mueq}) has been used with confidence to fit the rotation curves of
an impressive list of external galaxies with remarkable accuracy (Sanders
\& McGaugh 2002). However, very little work has
been carried out to study the actual effect of the curl field in the
non-relativistic MOND theory (Brada \& Milgrom 1995, 1999;
Ciotti, Londrillo \& Nipoti 2006), while the quantitative difference
between the predictions of MOND and TeVeS has not been studied at all. The
issue has become urgent as recent evidence against MOND is largely based on
multi-centred systems such as satellites of the Milky Way (Zhao 2005)
and the bullet cluster of merging galaxies (Clowe et al. 2004).

In this paper, after recalling some basic formulae of MOND and TeVeS
(\S\ref{sec:montev}), we
propose a set of parametric interpolating functions that are physical
in TeVeS (\S \ref{sec:inter}). Then we present spherical analytical
potential-density pairs for the baryon
distribution in early type galaxies, galactic bulges and dwarf
spheroidals, valid in MOND as well as in TeVeS (\S \ref{sec:spherpot}). For this model we work
out the density, potential, circular velocity, isotropic and anisotropic distribution functions.
We also show a simple analytical expression for the gravitational bending angle. 
Then we present the combination of such models in 
multi-centred systems for the classical MOND approximation(\S
\ref{sec:twocenter}); this provides an
indication of the kind of result we could expect from a rigorous modelling
of the bullet cluster of galaxies (Clowe et al. 2004). More
generally, we then show how to extend those spherical models to oblate ones
(\S \ref{sec:flatmond}). In those non-spherical situations, the density corresponding to a given
potential differs in the classical MOND and in TeVeS. To illustrate this,  we study the
effects of the flattening of the potential in the special case
of scale-free oblate one-dimensional models (\S \ref{sec:scalefree}), and we
explicitly show the subtle
differences between the different theoretical frameworks. Note that
Milgrom (1994) has also suggested that MOND could have a modified inertia basis
rather than a modified gravity basis; in that case the original MOND
formula is correct for circular orbits in galaxies, the theory would
become strongly non-local, the conservation laws would become unusual, and the
potential-density approach used hereafter would not apply to that framework. 

\section{MOND and TeVeS}\label{sec:montev}

In the aquadratic Lagrangian theory of MOND by Bekenstein \& Milgrom\
(1984), the Poisson equation reads
\begin{equation}
\protect\label{eqn:modpoi}
\grad . [ \mu \grad \Phi] = \grad^2 \Phi_N = 4 \pi G \rho,
\end{equation}
where $\mu(|\grad \Phi| / a_0)$ is the same interpolating function as in
Eq.(\ref{eqn:mueq}). We then have
\begin{equation}
\protect\label{eqn:mulast}
        {\mu}(g/a_0) \, {\bf g} = {\bf g}_{{\rm N}} + \grad \times {\bf H}.
\end{equation}
The value of the curl field depends on the boundary conditions, but
vanishes in spherical symmetry where Gauss' theorem applies and Eq.(\ref{eqn:mueq})
is recovered. In realistic geometries, the curl field is non-zero but
small (see Brada \& Milgrom 1995, 1999; Ciotti et al. 2006), leading to
small differences when computing the rotation curves of spiral galaxies.

On the other hand, Bekenstein's relativistic MOND (Bekenstein 2004) is a tensor-vector-scalar
(TeVeS) theory: the tensor is an
Einstein metric $g_{\alpha \beta}$ out of which is built the usual Einstein-Hilbert action. 
$U_\alpha$ is a dynamical normalized vector field ($g^{\alpha
  \beta}U_\alpha U_\beta = -1$), and $\phi$ is a dynamical scalar field. The
action is the sum of the Einstein-Hilbert action for the tensor $g_{\alpha
  \beta}$, the matter action, the action of the vector field $U_\alpha$,
and the action of the scalar field $\phi$. Einstein-like equations are
obtained for each of these fields by varying the action w.r.t. each of
them.

In TeVeS, the physical metric near a quasi-static
galaxy is given by the same metric as in General
Relativity, with the Newtonian potential $\Phi_N$ replaced by the total potential
\begin{equation}
\protect\label{eqn:xi}
\Phi = \Xi \Phi_N + \phi,
\end{equation}
where $\Xi \simeq 1$. This means that the scalar field $\phi$ plays the
role of the dark matter gravitational potential. The Einstein-like equation for the
scalar field relates it to the Newtonian potential $\Phi_N$ (generated by the baryonic density $\rho$) through the equation
\begin{equation}
\protect\label{eqn:poisson}
\grad . [ \mu_s \grad \phi] = \grad^2 \Phi_N = 4 \pi G \rho,
\end{equation}
where $\mu_s$ is a function of the scalar field strength $g_s=|\grad
\phi|$, and derives from a free function in the action of
the scalar field.

\section{The interpolating functions}
\protect\label{sec:inter}

In spherical symmetry, we have
\begin{equation}
\protect\label{eqn:mus1}
\mu_s g_s = \mu (g_s + g_N) = g_N,
\end{equation}
where $\mu$ is the interpolating function of MOND, thus related to $\mu_s$ by
\begin{equation}\label{eqn:mus}
\mu_s = \frac{\mu}{1-\mu}. 
\end{equation}
The standard interpolating function that has been used for twenty years to
fit the rotation curves with Eq.(1) is
\beq
\mu(x) = {x \over \sqrt{1+x^2}}.
\eeq
However, Zhao \& Famaey (2006) have shown that this function, or rather
the corresponding function $\mu_s$ derived from Eq.(\ref{eqn:mus}), is not
physical in the framework of TeVeS; the function $\mu_s$ is multi-valued. 
For this reason, we will use
hereafter a parametric set of interpolating functions that are physical in
TeVeS:
\beq
\protect\label{eqn:mux}
\mu(x) = {2x \over 1+ (2-\alpha) x + \sqrt{(1-\alpha x)^2 + 4x}}, ~ 0 \leq
\alpha \leq 1,
\eeq 
where $x = g/a_0$.
The corresponding scalar field functions are
\beq
\protect\label{eqn:musalph}
\mu_s(s) = {s \over 1-\alpha s},
\eeq
where $s = g_s/a_0$. The $\alpha=0$ case corresponds to the toy model
proposed by Bekenstein (2004) in weak and intermediate gravity.  Under
the approximation of Eq.(1), the $\alpha=1$ model
has been shown by Famaey \& Binney (2005) and Zhao \& Famaey (2006) to be a better fit to the rotation curves of galaxies than the $\alpha=0$ case. The values $0<\alpha<1$ have
not yet been explored in real galaxies, and will be the subject of another
paper (Famaey, Gentile \& Zhao 2006, in preparation). 

These functions $\mu$ and $\mu_s$ will lead to the same gravitational behaviour
in spherical symmetry. However, note that Eq.(\ref{eqn:mus}) is not valid in a more general geometry: the Newtonian
force, the MOND force of Bekenstein \& Milgrom (1984) and the TeVeS force
are no longer parallel. The curl field obtained when solving the equation
for the scalar field $\phi$ (Eq. \ref{eqn:poisson}) will be different than the one
obtained when solving for the full $\Phi$ in Eq.(\ref{eqn:modpoi}). This
will be illustrated in \S 7.

\section{Spherical potential-density pairs in MOND/TeVeS}
\protect\label{sec:spherpot}

In this section, we consider only the $\alpha=1$ case, known to yield
excellent fits to the rotation curves of galaxies. Potential-density pairs (e.g., Hernquist 1990, Dehnen 1993, Zhao 1995)
have long been acknowledged to be very useful in model building and
in checking numerical simulations. It is
even more interesting to find simple spherical and
non-spherical models in MOND and TeVeS.
We explore a spherical model with a scalar field of the form
\begin{equation}
\phi(r) = v_0^2 \ln \left(1+ {r \over (p+1) r_0} \right), 
\end{equation}
where $p$ is a dimensionless number, and
\begin{equation}
\protect\label{eqn:M}
r_0 \equiv {v_0^2 \over a_0},~v_0^2=\sqrt{G M_0 a_0}.
\end{equation}
Then we have, according to the equation for the scalar
field (Eq. \ref{eqn:poisson}), that
\begin{equation}
\protect\label{eqn:rho2}
\rho(r) = {M_0 \left[ r_h (r + r_h) - r_0^2/4 \right] \over 2\pi r \left[ (r+r_h)^2 - r_0^2/4 \right]^2 }, \qquad r_h=(p+{1 \over 2})r_0
\end{equation}
\begin{figure}
\includegraphics[angle=0,width=8.5cm]{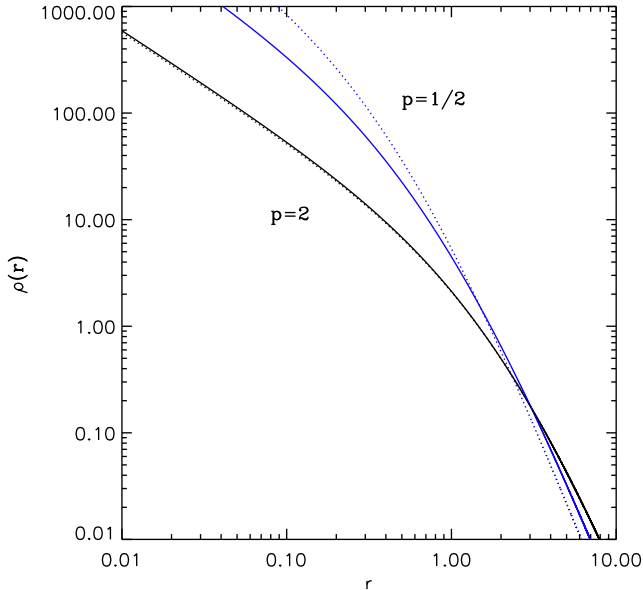}
\caption{Shows the density of Eq.(\ref{eqn:rho2}) for p=0.5 and 2.0. $r_0$ and $v_0^2$ are taken to be unity, and G=0.00442 in all following plots unless stated otherwise.   Overplotted is a Hernquist density profile (cf. Eq. \ref{eqn:hern}) for p=0.5 and 2.0. It shadows the curve for p=2 from
our density profile extremely well, but there is a slight deviation between the p=0.5 curves.}
\protect\label{fig:rhos}
\end{figure}
The density profile is shown for p=0.5 and 2.0 in Fig.1. This MONDian density distribution is realistic: the model mimics the Sersic
profile of an elliptical galaxy. It also
mimics a Hernquist (1990) model with scalelength $r_h$: 
\begin{equation}\protect\label{eqn:hern}
\rho_h(r) = {M_0 r_h \over 2\pi r (r+r_h)^3}.
\end{equation}
We can then also easily derive $g_s = v_0^2/(r+r_0+pr_0)$, and from Eq.(\ref{eqn:mus1})
calculate the Newtonian and total potential of the model:
\beq
\Phi_N(r) = \Phi - \phi(r), \qquad \Phi (r) = v_0^2 \ln \left(1+ {r \over p r_0 }\right).
\protect\label{eqn:phirho}
\eeq
Thus, for the gravity we have
\begin{equation}
\protect\label{eqn:gofr}
g(r) = {V_{\rm c}^2 \over r} = {v_0^2 \over r+ p r_0}.
\end{equation}
The corresponding matter density (baryons + DM) in the Newtonian framework is
\beq\label{eq:density}
\rho(r) + \rho_{DM}(r)= {M_0 (1 + {2pr_0 \over r})\over 4\pi r_0  (pr_0+r)^2 }.
\eeq
The asymptotic circular velocity is $v_0$, and the maximum gravity $g_0$
occurs at small r (near the origin), where $g(r) = {v_0^2 \over p r_0} = {a_0 \over p}$ (note that in the flattened models
 of \S 7 $g_0$ will be related to p by $g_0={a_0 \over p}$).
Given this, it is interesting to show the spherical model
with p=0.5 and 2.0, hence the intermediate gravity range $g_0=2a_0$ and
$a_0/2$ (Figs. 1-4).
Most bulges and ellipticals, gravitational lenses and galaxy clusters
should go through this intermediate regime,
while dwarf spheroidal galaxies have very low $g_0/a_0={1 \over p}$. Although a detailed fit to observed light profile is beyond the scope of the
present paper, we will show an application in galaxy clusters in \S 5.  

\subsection{The baryonic distribution function}
\protect\label{sec:bardist}

It is interesting to ask if 
the MONDian potential-density pairs 
presented in the previous section can be realised
by some equilibrium configurations
described by certain baryonic distribution functions self-consistently.
This can be done in exactly the same way as in Newtonian gravity.  
This is particularly simple for a system with constant radial anisotropy 
such that the radial velocity dispersion $\sigma_r$ is related to 
the tangential dispersions by
$\sigma_r^2 = 2\sigma_{\theta}^2 = 2\sigma_{\phi}^2$.  
In this case the baryonic phase-space distribution $F$ for
the distribution function must take the form: 
\beq
F (E,L) = {A(E) \over 4 \pi L}, \qquad
A(E) \equiv -{d (r\rho) \over d \Phi }\biggr|_{\Phi=E}.
\eeq
Knowing from Eq.(\ref{eqn:phirho}) that
\beq
r(\Phi)= p r_0\left(\exp{\Phi \over v_0^2} -1\right),
\eeq
we obtain
\begin{equation}\label{eqn:anisdf}
F(E,L) = {M_0 \over 16\pi^2 r_0^2 v_0^2 L}{\Lambda^3\left(2\Lambda^2 + (6p+1)\Lambda + 3p(2p+1)\right) \over p(p+\Lambda)^3},\qquad \Lambda = \exp{-E \over v_0^2}.
\eeq

For this radially anisotropic model we can 
calculate the radial velocity dispersion by solving the Jeans equation
as done for Newtonian system (e.g., Angus \& Zhao 2006) 
\beq
2\sigma_{\theta}^2 = 2\sigma_{\phi}^2=\sigma_r^2(r)
={1 \over r \rho} \int_r^\infty V_{\rm c}^2 \rho dr.
\eeq
Substitute in the expression for the circular velocity $V_c$ we have
\beq
\sigma_r^2(r) =
v_0^2 {(s^2-1)^2 \over 4(2p+1)s-4}
\left[ ({p \over 2}+1) \left( \ln {s-1 \over s+1} +{2 \over s+1} \right)
+ {2 \over s^2-1} + {p \over (s-1)^2} \right]
\protect\label{eqn:sigr}
\eeq
where $s \equiv 1+ 2 p + {2 r \over r_0}$.  A likewise expression can be
found by calculating the moments from the distribution function $F(E,L)$.

An interesting feature of this model is that it has {\it nearly
isothermal} velocity dispersions everywhere for any value of the
parameter $p$.  Fig.(\ref{fig:asig1}) shows $\sigma_r(r)$ for the two
limiting cases of $p=0$ and $p=\infty$.  To understand this note that
at large radii $\rho \sim r^{-4}$, $V_{\rm c} \rightarrow v_0$ and
$\sigma_r^2 \rightarrow {v_0^2 \over 3}$.  At small radii, $r \rho
\rightarrow cst$, and $\sigma_r^2(0) = (p+1)v_0^2\left[ 3/4+p/2
+ p(p/2+1) \ln{p \over p+1} \right] = 
0.75v_0^2,~0.470v_0^2,0.421v_0^2,0.383v_0^2,0.333v_0^2$ for 
$p=0,0.5,1,2,\infty$.  So for all plausible values of p,
$\sigma_r^2$ is very comparable at the center and at large radii.
The anisotropic distribution function as a function of $E$ is 
presented in Fig.(\ref{fig:anisdfig}) for p = 0.5 and p= 2.0.
We can also numerically integrate the isotropic distribution function via
Eddington's equation (see appendix and Fig.\ref{fig:isodf}).

\begin{figure}
\includegraphics[angle=0,width=8.5cm]{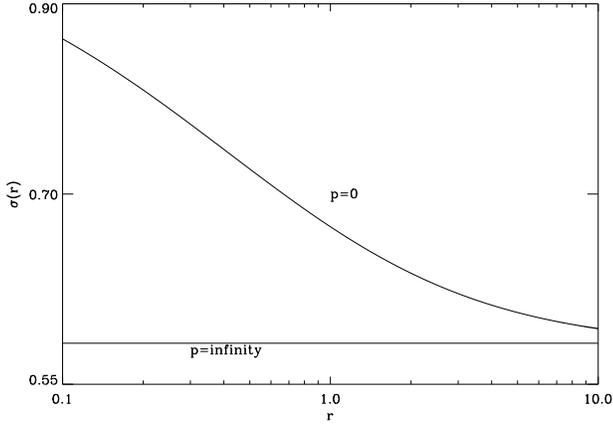}
\caption{Shows the radial velocity dispersion of anistropic model given by Eq.(\ref{eqn:sigr}) for the limiting cases p=0 and p=$\infty$. Here $v_0$ and $r_0$ are unity.}
\protect\label{fig:asig1}
\end{figure}

\begin{figure}
\includegraphics[angle=0,width=8.5cm]{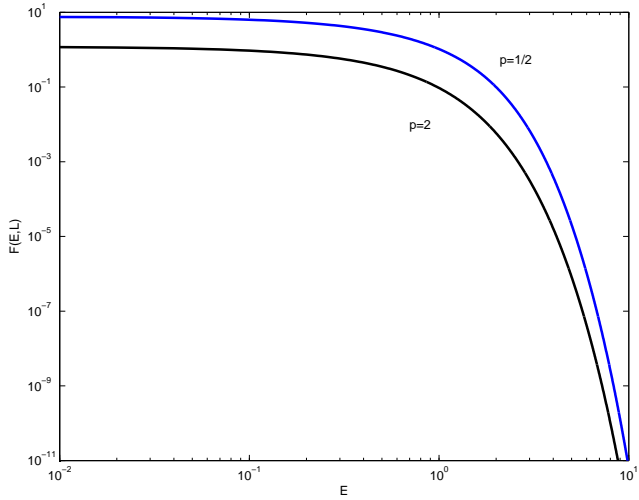}
\caption{Shows the anisotropic distribution function from Eq.(\ref{eqn:anisdf}) at
  $L=1$ for p=0.5 and 2.0}
\protect\label{fig:anisdfig}
\end{figure}

\begin{figure}
\includegraphics[angle=0,width=8.5cm]{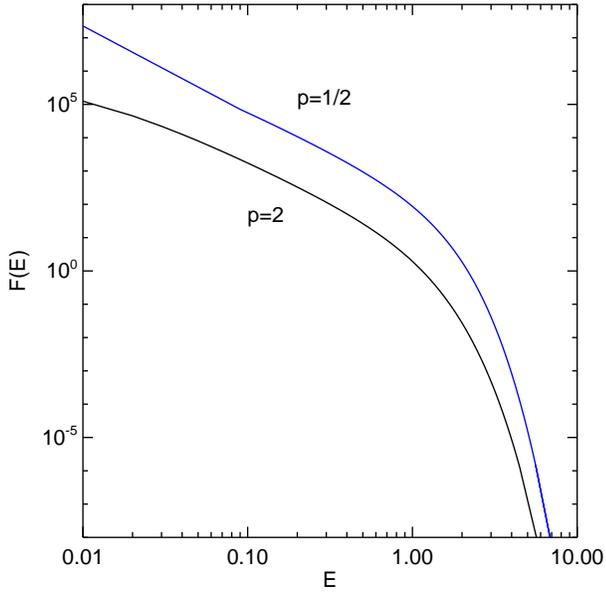}
\caption{Shows the isotropic distribution function derived in the appendix for p=0.5 and 2.0. }
\protect\label{fig:isodf}
\end{figure}

\begin{figure}
\includegraphics[angle=0,width=8.5cm]{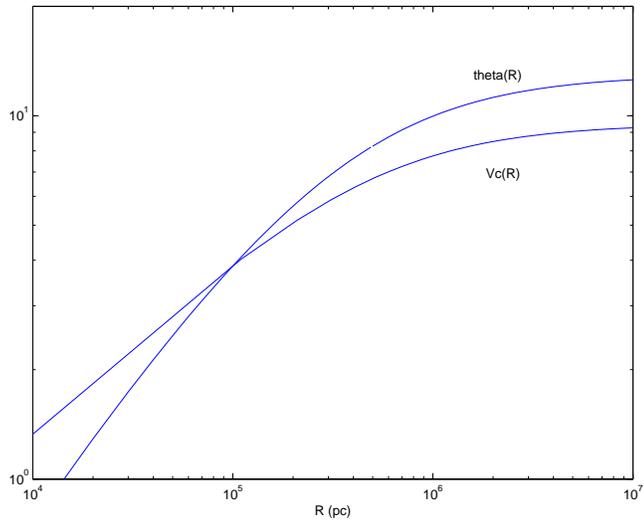}
\caption{Shows the bending angle $\theta$ in arcseconds as a function of impact parameter R and the circular speed $V_c$ in $10^2$$\kms$ for $pr_0$=0.5Mpc and $v_0$=949$\kms$.}
\protect\label{fig:ben}
\end{figure}

\begin{figure}
\includegraphics[angle=0,width=8.5cm]{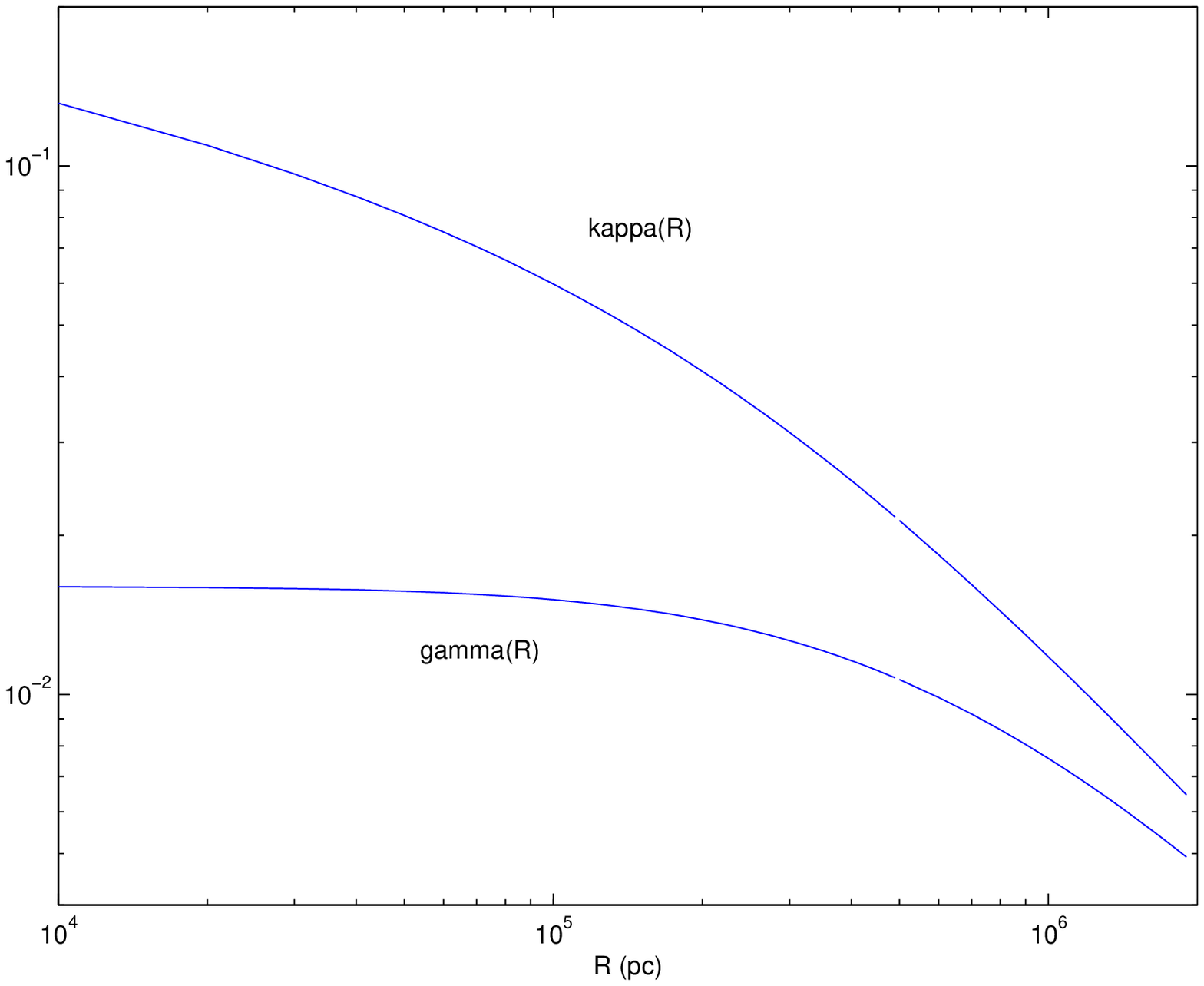}
\caption{Shows the convergence $\kappa$ and tangential shear $\gamma$ for a
  toy cluster at an effective distance $D_{\rm eff}$=400Mpc with $v_0$=949$\kms$ and p$r_0$=0.5Mpc.}
\protect\label{fig:kappa}
\end{figure}

\subsection{Gravitational lensing}
\protect\label{sec:gravlens}

Light bending in TeVeS works very much like in General Relativity.  For a ray of impact
parameter $R$ from the spherical lens,  the bending angle $\theta$ is
given (cf. Zhao, Bacon, Taylor, Horne 2006) by the following integration
along the line of sight
\beq
\theta (R)  = \int_{-\infty}^{+\infty} {2 g_\perp dz \over c^2},
\qquad g_\perp (R,z) = g(r) {R \over r},
\eeq
where $g_\perp (R,z)$
is the gravity perpendicular to the line of sight, and $g(r)$ is the
centripetal gravity at radius $r=\sqrt{R^2+z^2}$.  Using the
expression for $g(r)$ from our model (Eq. \ref{eqn:gofr}), we find that
the deflection angle is given by
\beq
\label{eq:ben}
\theta (R) = 
\cases{
{8 v_0^2 \over c^2 \sqrt{1-y^{-2}}} \arctan \sqrt{y-1 \over y+1} & when $ y\equiv {R  \over p r_0} > 1$ \cr
{4v_0^2 \over c^2} & when $y\equiv {R  \over p r_0}=1$\cr
{8 v_0^2 \over c^2 \sqrt{y^{-2}-1}} \tanh^{-1} \sqrt{1-y \over y+1} & when $y\equiv {R  \over p r_0} < 1$ \cr
}
\eeq
Fig.\ref{fig:ben} shows the predicted bending
angle as function of the impact parameter $R$; cases are shown for $p=0.5$
and $2.0$. The bending angle increases with the impact parameter and starts to level off beyond $R=p r_0$.  This is consistent with the expectation of
a flat rotation curve at large radii (Fig. \ref{fig:ben}).
Given the distance to the lens $D_l$, the distance to the source $D_s$ and
the lens-source distance $D_{ls}$ we can define an effective distance as
$D_{\rm eff}=D_l D_{ls}/D_s$. Using this we can compute the convergence $\kappa$ and the tangential shear $\gamma$ from
\bey\label{eq:kappasph}
\nonumber
\kappa(R) &=& {\theta D_{\rm eff}  \over R} - \gamma(R)\\
&=&{R^2-2(pr_0)^2\over 2 R^2-2(pr_0)^2}{(\theta -\theta(pr_0))D_{\rm eff}\over R},
\eey
and they are plotted for a toy cluster in Fig.(\ref{fig:kappa}).

\section{Multi-centred potentials and the merging bullet cluster}
\protect\label{sec:twocenter}

\begin{figure}
\includegraphics[angle=0,width=8.5cm]{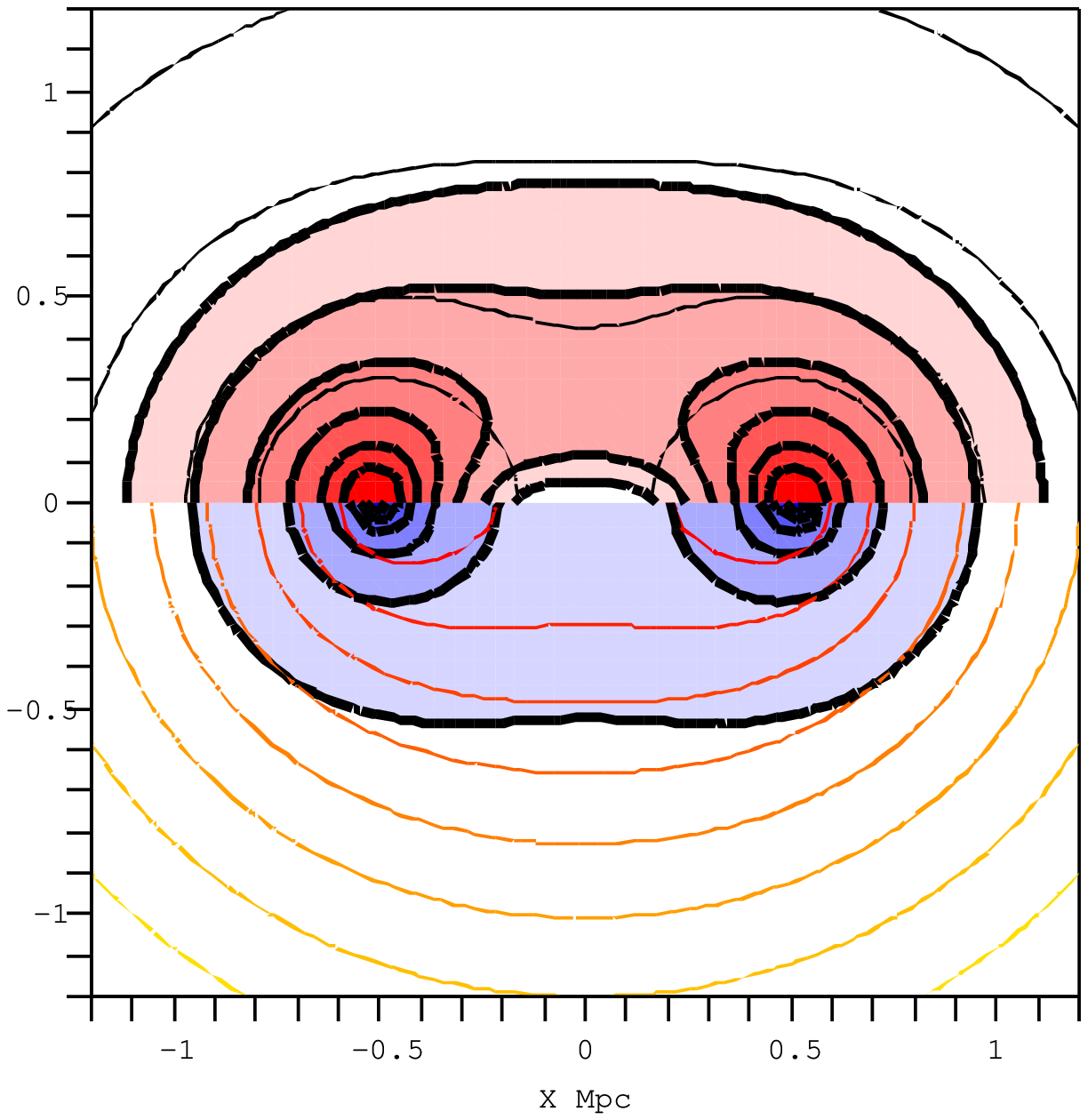}
\includegraphics[angle=0,width=8.5cm]{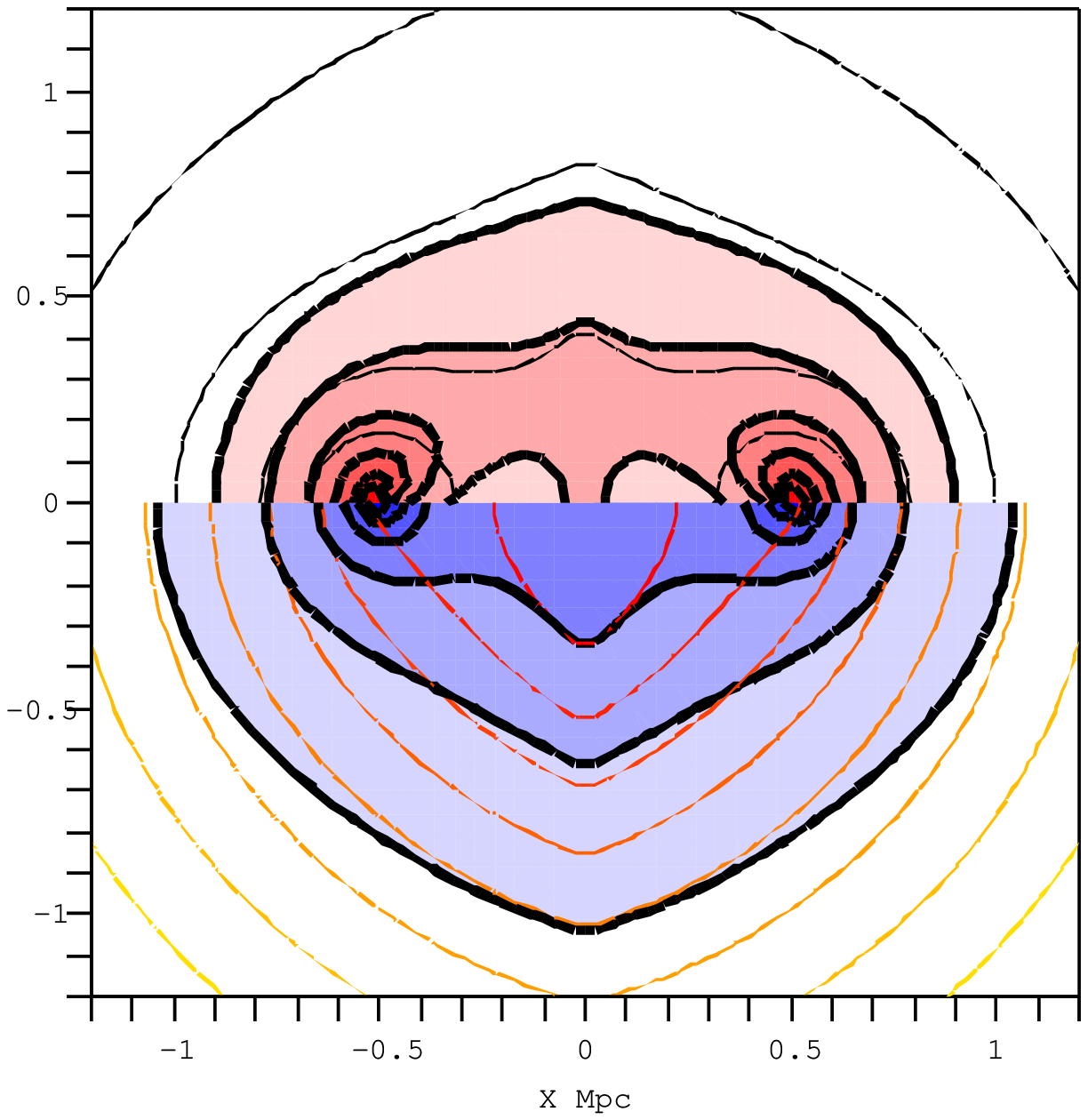}
\caption{Shows contours of a double-centred (panel a) and 
a triple-centred (panel b) analytical model for the
potential $\Phi$ (dashed red lower) and 
projected lensing convergence $\kappa$ (thick blue shaded lower) 
of a system resembling the bullet cluster  (Clowe et al. 2004).  
Also shown are contours 
(spaced in 0.5 dex, in internals of factor of two)
the matter volume density in Einstein-Newton gravity (upper thin black) 
and baryonic matter volume density in the classical MOND approximation
  (upper shaded zones).  
Note a density minimum for 'MONDian' baryons at (0,0) in panel (a), and 
a density maximum in panel (b) 
where the baryons are primarily in a disky component
in the middle. }
\protect\label{fig:bullet}
\end{figure}

An obvious way of obtaining flattened models is to consider a
double-centred potential, whose axis of symmetry is 
perpendicular to the line of sight, i.e., a merging system viewed edge-on.  
In a cartesian coordinate system $(x,y,z)$, if
the two centres are located at $(-x_0,0,0)$ and $(x_0,0,0)$, then the
double-centred potential can be written as
\beq
\Phi_{\rm dc}(x,y,z) = { \Phi(r_1) \over 2 } + { \Phi(r_2) \over 2 },  
\eeq
where $r_1 = \sqrt{(x+x_0)^2 + y^2 + z^2}$, $r_2 = \sqrt{(x-x_0)^2 +
 y^2 + z^2}$, and $\Phi$ is the spherical potential of Eq.(16). 
Here we have simply added two spherical analytical potentials.  
For example, to mimic a merging galaxy cluster we set $p=2$, $x_0 = p r_0 =
0.5$Mpc, $v_0 = 949$ km/s, and $M_0=5.21 \times 10^{13}\msun$. 

The gravitational field of such a model is easily computed by
the simple superposition of two spherical fields, i.e., 
\beq
{\bf g} =
 - {v_0^2 \over r_1+ p r_0} {{\bf r}_1 \over 2r_1}
 - {v_0^2 \over r_2+ p r_0} {{\bf r}_2 \over 2r_2}.
\eeq
Likewise the light bending angle vector is the superposition of 
the bending angle vectors of two spherical models, 
so its $x$ and $y$ direction components are respectively 
\beq
\left \{
\begin{array}{c @{ =} c}
\theta_{{\rm dc} x} (x,y) & \frac{(x+x_0)\theta(R_1)}{2R_1} +
\frac{(x-x_0)\theta(R_2)}{2R_2}\\ 
\theta_{{\rm dc} y} (x,y) & \frac{y\theta(R_1)}{2R_1} +
\frac{y\theta(R_2)}{2R_2}, \\
\end{array}
\right.
\eeq
where $R_1 = \sqrt{(x+x_0)^2 + y^2}$, $R_2 = \sqrt{(x-x_0)^2 + y^2}$, and $\theta$ is the spherical bending angle of Eq.(\ref{eq:ben}). From
there we can estimate the projected lensing convergence
\beq
\kappa(x,y) = 
{1 \over 2}D_{\rm eff} \left( \frac{\partial \theta_{{\rm dc}
    x}}{\partial x} + \frac{\partial \theta_{{\rm dc}
    y}}{\partial y} \right) 
= {\kappa(R_1) \over 2} + {\kappa(R_2) \over 2},
\eeq
where $\kappa(R)$ is given by Eq.(\ref{eq:kappasph}) 
for spherical systems.  
Assuming $D_{\rm eff} = 400$Mpc, 
the resulting convergence contours
are plotted on Fig. \ref{fig:bullet}(a).  This map resembles the 
convergence map derived from the weak lensing shear field around 
the merging bullet cluster 1E 0657+56 (Clowe et al. 2004). 

So far the result does not 
differ from Einsteinian gravity. The difference is in the 
underlying matter distribution.
In Einsteinian gravity, the convergence map is simply proportional
to the surface density map, and the volume density of matter is 
the simple addition of two spherical models each with density given 
in Eq.(\ref{eq:density}).  As we can see from Fig.\ref{fig:bullet} 
there is a one-to-one relation between features in 
the convergence map and the features in the matter (dark plus baryons) 
distribution if the gravity is Einsteinian, i.e., what we see in lensing
is what we have.

The situation in MOND/TeVeS is different. In order to properly derive the
corresponding density in TeVeS, one would need to know what part of the
double-centred potential is due to the scalar field. This will be done in the limiting case of scale-free flattened
models in \S 7. Here we use the classical MOND approximation. From the
gravity or the potential we can then directly calculate the corresponding
baryonic isodensity contours using Eq.(2) and Eq.(9) with $\alpha=1$.
The MONDian baryonic matter has a rich structure.  
Fig.\ref{fig:bullet}(a) shows that at the centre $(x=0,y=0)$ 
the MONDian volume density 
reaches a local minimum while the convergence map shows a saddle point.  
This cautions against a naive deprojection of the convergence map in
MOND/TeVeS. 

\subsection{Triple-centred baryonic system and the bullet cluster 1E 0657+56}
In the case of the bullet cluster, there are three baryonic mass 
concentrations. Clowe et al. (2004) argue that 
the bulk of baryonic mass is in the form of X-ray gas, which is shocked
and displaced from the two optical centres of the colliding binary cluster.  
It was argued that lensing in any MONDian theory should produce shear maps 
centred on the dominating X-ray gas instead of the lesser baryonic mass
responsible for the optical light.
The fact that the convergence map coincides with the two 
optical centres is presented as direct evidence for the presence of
collisionless dark matter, unaffected by the shock, and respecting the
optical centres.   

Of course, this is not so surprising since it is well known that MOND still
needs an unseen matter component in galaxy clusters (Sanders 2003). But, in
the case of the bullet, a key element of the line of reasoning is
that the geometry of the lensing map in TeVeS reflects the underlying
baryons even in highly non-spherical geometries.  To illustrate the range
of possibilities in triple-centred systems, let us consider the following
potential  
\beq
\Phi_{\rm tc}(x,y,z) = 
\left[k_1 + (1-k_1-k_2) H(x) \right] \Phi(r_1) 
+ \left[k_2+ (1-k_1-k_2) H(-x)\right] \Phi(r_2).
\eeq
The terms involving the Heaviside-function create the effect
of a razor thin disk at $x=0$ (reminiscent of the well-known Kuzmin disk),
with  a sudden change of gravity in the $x$-direction at the midplane $x=0$.  
The potential is continuous across the $x=0$ plane. The deflection
angle and convergence map remain those of the superposition of 
two spherical deflectors, although there is now a sudden change 
in the weighting of the two deflectors
as one crosses the midplane $x=0$.  Finally one can apply 
the MONDian Poisson equation (Eq. 2) to derive the baryonic density.

Here we choose $k_1=k_2=0.2$ so that we have a prominent baryonic 
component in between the two cluster centres.   Fig.\ref{fig:bullet}(b)
shows a rather regular looking convergence map, but a complex 
MONDian baryonic density distribution, 
somewhat resembling that of the bullet cluster.
In particular, there is now a ridge of matter centred on 
$x=0$ and two {\it irregular} baryonic components
more or less centred on the $x=\pm x_0$.  This example provides evidence that
a regular-looking convergence map is not incompatible with 
a MONDian multi-centred baryonic mass distribution.

As argued in Zhao, Bacon, Taylor \& Horne (2006) and Zhao \& Qin (2006)
the convergence $\kappa$ can be {\it non-zero} where there is no projected
matter in MOND/TeVeS, something that is not possible in Einsteinian gravity.  This implies that a lensing convergence 
map does not simply translate a baryonic surface density map 
in MOND.  Our models here show the range of non-trivial 
baryonic geometries for multi-centred potentials in MOND.
Although the models here are of only indicative value (as
the approximation of classical MOND has been used, see \S 7 for typical
differences with TeVeS) it does caution
drawing conclusions about TeVeS before performing a careful lensing
analysis of the bullet cluster (Clowe et al. 2004) in the framework of
TeVeS. 

\section{Flattened potentials and orbits}  
\protect\label{sec:flatmond} 
The spherical potential of \S 4 can also be generalised
into a flattened or triaxial potential. For example, we can consider
a model with an axisymmetric potential
\begin{equation}
\protect\label{eqn:phirthet}
\Phi(r,\theta) = {v_0^2 \over 2} \ln \left[ \left(1+ {r \over
    pr_0}\right)^2 + 2 \epsilon{r \over pr_0} \cos^2 \theta  \right] 
\end{equation}
where $\theta$ is the angle with the $z$-axis:
$\epsilon >0$ corresponds to an oblate potential and $\epsilon <0$ to a
prolate potential.

The corresponding density in the classical MOND of Bekenstein \& Milgrom
(1984) can be calculated by
feeding this potential into the modified Poisson equation (Eq. \ref{eqn:modpoi}) in spherical coordinates:
\beq
4\pi G \rho(r,\theta) =
{\partial \over r^2 \partial r} {\mu \partial \Phi \over \partial r}
+ {\partial \over r^2 \sin \theta \partial \theta } {\mu \sin \theta
\partial \Phi \over \partial \theta}.
\eeq
The expression for $\rho$ can be obtained analytically,
but the general expression is too tedious to be given here.
Nevertheless we can calculate orbits in this potential numerically.
The stars in this flattened MOND potential are typically on loop orbits
as in a flattened Newtonian potential (Fig. \ref{fig:looporbits}).  We
shall concentrate hereafter on a limiting case of this flattened model, in
order to compare the predictions of the classical MOND gravity and of TeVeS. 

\section{Scale-free flattened models: multi-field theory vs. one-field theory}
\protect\label{sec:scalefree}

\begin{figure}
\includegraphics[angle=0,width=8.5cm]{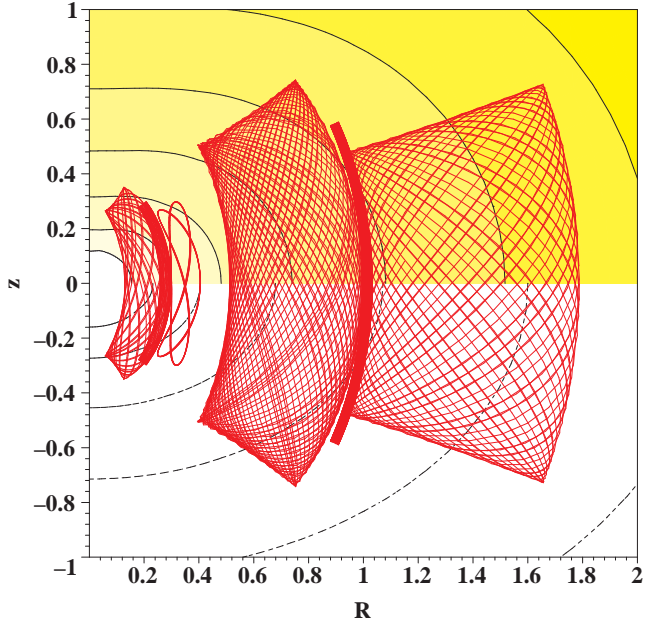}
\caption{Shows a flattened non-scale-free potential with $g_0=a_0/2$ (p=2), $\epsilon$=0.25,
upper panel shading showing density contours (for $\alpha=1$),
bottom dashed showing Hernquist density profile (cf. Eq. \ref{eqn:hern}) with axis ratio 2/3.}
\protect\label{fig:looporbits}
\end{figure}

In the limit $r/pr_0 \rightarrow 0$, the previous potential reduces to a
scale-free form when expanding Eq.(\ref{eqn:phirthet}). We shall now consider such models
with a total potential 
\beq\label{eqn:phirth}
\Phi(r,\theta) = r g_r(\cos\theta),~{\rm where} ~ g_r(c) =  (1 +
\epsilon  c^2)g_0 , 
\eeq
and for convenience we define
\beq
c \equiv cos \theta.
\eeq
Clearly, $g_0=v_0^2/pr_0$ is thus the gravity along the major axis in the
$c=0$ (or $\theta = \pi/2$) plane, and the parameter $\epsilon$ leads to
flattening ($\epsilon=0$ reduces to the spherical case). The parameter
$g_0$ is NOT the acceleration constant $a_0$ of MOND, but is a parameter of
the model linked to it through the definitions (see Eq. \ref{eqn:M}) of $v_0$ and $r_0$
($g_0=a_0/p$). 

This class of scale-free models corresponds to flattened
galactic bulges, ellipticals, dwarf spheroidal galaxies, or centre of
galaxy clusters 
depending on the value of the equatorial gravity $g_0$. The
radius-independent gravitational force in these models is: 
\beq
{\bf g}(c) = \grad \Phi(r,\theta) = g_r(c) \hat{r} + g_\theta(c)
\hat{\theta},
\eeq
\beq
{\rm where} ~
g_\theta(c) = (1-c^2)^{1/2}\left({dg_r \over dc}\right) = 2 c \epsilon g_0
(1-c^2)^{1/2}.
\eeq
The amplitude of gravity $g$ is defined as
\beq\label{g}
g(c)=[g_r(c)^2+g_\theta(c)^2]^{1/2}.
\eeq 

We are now going to explore {\it the density corresponding to this potential in
four different frameworks:} (i) Newtonian gravity where the computed density
corresponds to the baryonic and DM distribution, (ii) classical
one-field MOND gravity (Bekenstein \& Milgrom 1984, approximation used in
\S 5), (iii) TeVeS multi-field theory, (iv) original MOND formula
(Eq. \ref{eqn:mueq}) approximation.\footnote{In a flattened system, Eq.(1) does not derive
from a proper theory of gravity (energy is generally not conserved). The only way to precisely recover it is to
consider purely circular orbits in axisymmetric systems for non-relativistic
modified inertia toy models (see Milgrom 1994). In a flattened system, the
corresponding formula for the classical one-field MOND gravity will differ
from Eq.(1) by a curl-field term (Eq. \ref{eqn:mulast}), while this
curl-field will only affect the scalar field in TeVeS.}

For the first time in the literature, we provide here a quantitative
comparison of the three ways of describing the MOND phenomenology without
invoking DM (and of Newtonian gravity with DM), by
comparing the underlying density corresponding to the potential of
Eq.(\ref{eqn:phirth}).

\subsection{Newtonian gravity with dark matter}\label{sec:newtgrav}

Assuming Newtonian gravity, the density (corresponding to the stellar and
dark densities) is
\beq
\rho(r,\theta) = (4\pi G)^{-1} \nabla^2 \Phi(r,\theta) = (4\pi G r)^{-1}
L(c),
\eeq
\beq\label{L}
{\rm where} \qquad L(c) = 2 g_r(c) + {d\over dc} [(1-c^2)g_r'(c)]  = {\cal
  L} [g_r(c)] = 2 (1+\epsilon) g_0 - 4 \epsilon g_0  c^2 ,
\eeq
We have defined the differential operator ${\cal L}=2+{d \over dc}(1+c^2){d \over dc}$. This toy model
has a $r^{-1}$ cusp,  a rising rotation curve $V \propto \sqrt{r}$, and
$M(r) \propto r^2$ (similar to uniform disks).

\subsection{One-field MOND gravity}\label{sec:onemon}

Now, assuming a classical MOND gravity (as we did in \S 5) with an
interpolating function $\mu(g)$, the underlying (purely baryonic) density
can be computed from  
\beq
\rho_M(r,\theta) =(4\pi G)^{-1} \nabla.(\mu \nabla \Phi(r,\theta)) = (4\pi G r)^{-1} L_M(c),
\eeq
\beq
{\rm where} \qquad L_M(c) = \mu(g) L(c) + (1-c^2) {d \mu(g) \over dc}
{dg_r\over dc},    
\eeq 
where $L(c)$ is given by Eq.(\ref{L}) and $\mu(g)$ by Eq.(\ref{eqn:mux})
with $g$ as in Eq.(\ref{g}).
This baryonic density $\rho_M(r,\theta)$ is compared with the baryonic+DM
density of a Newtonian model with the same potential in
Fig.\ref{fig:galp}, for different values of $g_0$ and different
interpolating functions.

\subsection{Multi-field TeVeS gravity}\label{sec:multev}

To compute the baryonic density in \S 5 when making a toy model of the bullet
cluster, we used the one-field MOND gravity. We are
now going to show the differences that could be expected when using the
multi-field TeVeS instead. The problem in that case is that we must find the
relative contribution of the scalar field and of the Newtonian potential to
the total potential before computing the underlying baryonic density.

Assuming a multi-field gravity (see Eqs. 4 and 5) with a scalar
field $\phi$ and a Newtonian potential $\Phi_N = \Phi-\phi$, we have that
the scalar field can be written in the form
\beq
\phi(r,\theta) = r g_{s,r}(\cos\theta),~ {\rm where} ~ g_{s,r}(c) = g_r(c)
-g_{n,r}(c) = g_0(1+\epsilon c^2) - g_{n,r}(c),
\eeq 
where $g_{n,r}$ and $g_{s,r}$ are respectively the Newtonian gravity in the
$\hat{r}$ direction and the
scalar field strength in the $\hat{r}$ direction, both generated by the
common density distribution we are looking for,
$\rho_s(r,\theta)$.

The total scalar field strength is
\beq
g_s(c) = |\grad \phi(r,\theta)| = (g_{s,r}^2(c) + g_{s,\theta}^2(c))^{1/2}, ~ {\rm
  where} ~ g_{s,\theta}= (1-c^2)^{1/2}\left({dg_{s,r} \over dc}\right).
\eeq
The baryon density is then given by both the scalar and Newtonian Poisson
equations: 
\beq
\rho_s(r,\theta) = (4\pi G r)^{-1} \{\mu_s {\cal L}[g_{s,r}(c)]  + (1-c^2) {d\mu_s(g_s) \over dc} {dg_{s,r} \over dc} \},
\eeq
and
\beq
\rho_s(r,\theta) = (4\pi G r)^{-1} {\cal L}[g_{r}(c)-g_{s,r}(c)]
\eeq
Combining the above two equations, we use the numerical relaxation method
to solve the following second-order ODE for $g_{s,r}(c)$: 
\beq
[1+\mu_s] {\cal L}[g_{s,r}(c)]  + (1-c^2) {d\mu_s(g_s) \over dc} {dg_{s,r}
  \over dc} = L(c). 
\eeq
We set the boundary conditions such that the solution is regular at $c=\pm
1$. From this we derive the baryonic density $\rho_{s}(r,\theta)$.

We plot the equal density contours predicted from the three
theories for a model with  $g_0=a_0/2$ and $2a_0$, $\alpha$=0.0 and 1.0 and
$\epsilon$=0.8 in Fig.\ref{fig:galp}. Note that some of the models have
an unphysical heart-shaped density distribution. This doesn't prevent us
comparing them in order to show the differences between the different
formulations of MONDian modified gravity theories. 

\begin{figure*}

\def\subfigtopskip{0pt}    
\def\subfigbottomskip{4pt}
\def\subfigcapskip{1pt}
\centering

\begin{tabular}{cc}
\subfigure[]{\includegraphics[angle=0,width=8.5cm]{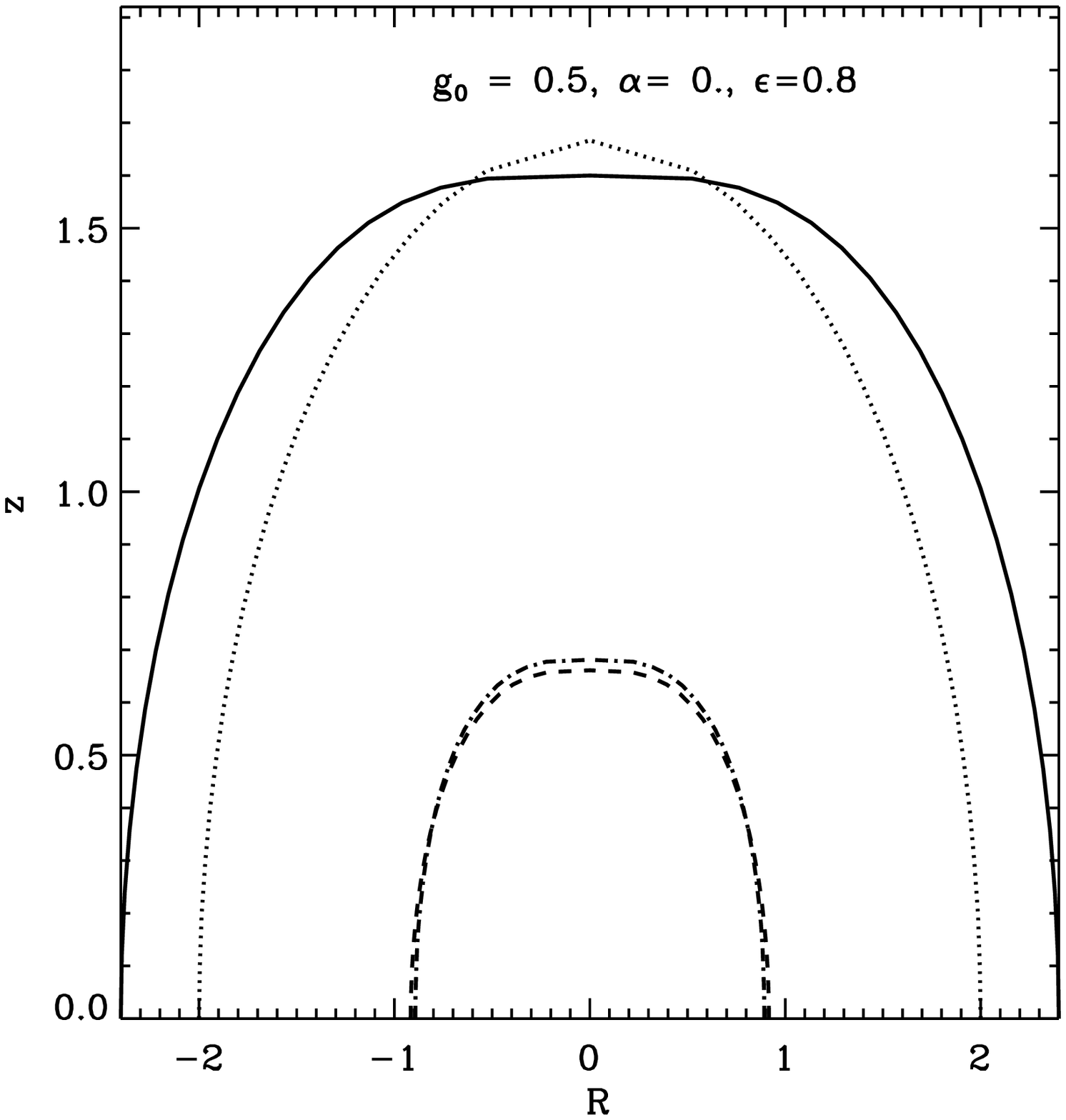}
}&
\subfigure[]{\includegraphics[angle=0,width=8.5cm]{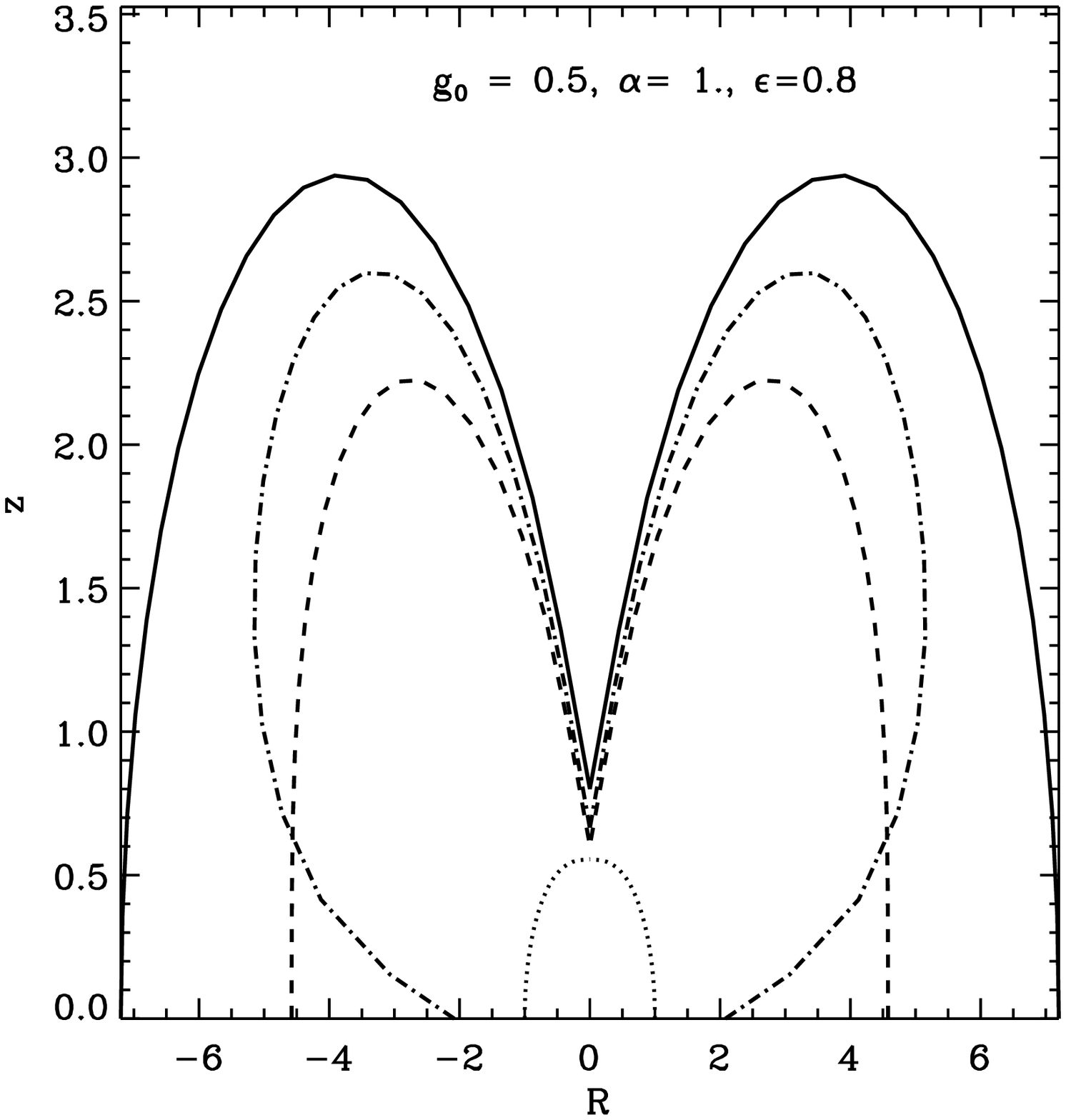}
}\\
\subfigure[]{\includegraphics[angle=0,width=8.5cm]{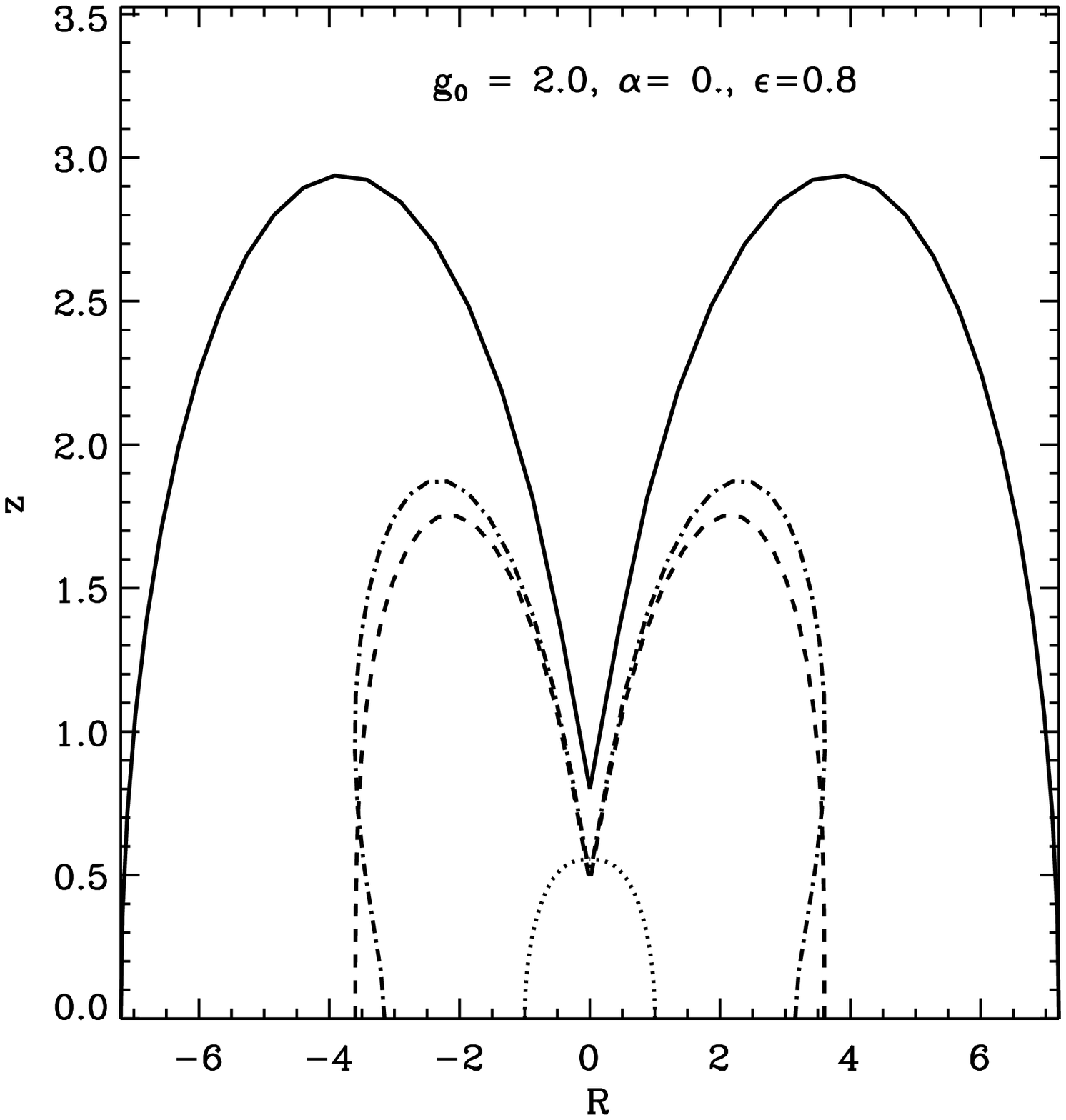}
}&
\subfigure[]{\includegraphics[angle=0,width=8.5cm]{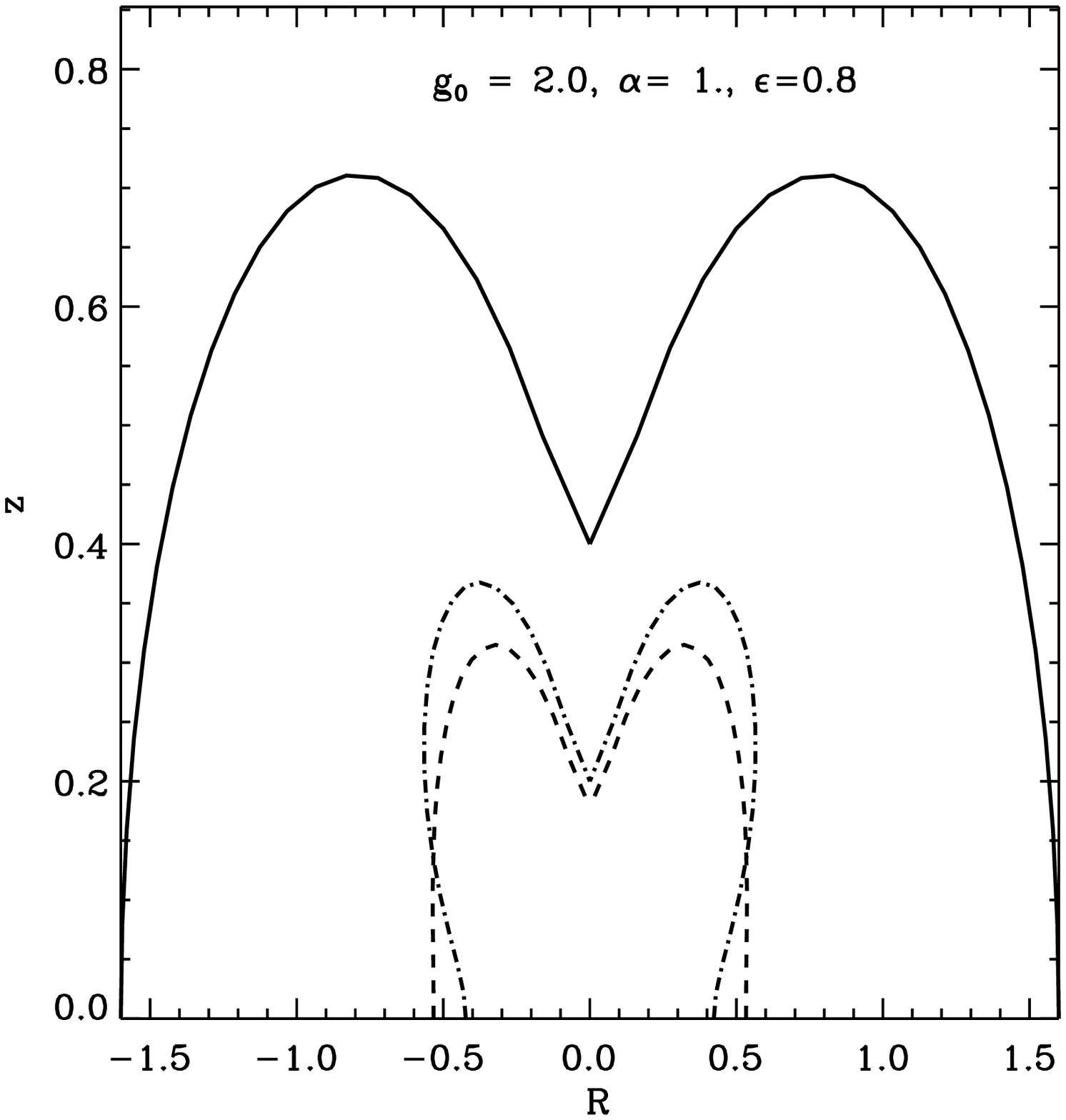}
}\\
\end{tabular}
\caption{Shows isopotentials (dotted line) for the potential of
  Eq.(\ref{eqn:phirth}), for 2 values of the equatorial gravity $g_0$ ($g_0=a_0/2$ and $g_0=2a_0$), and
  2 values of $\alpha$ (see Eq.(9)). The full black line corresponds to
  isodensity contours in Newtonian gravity (baryons+DM). The dashed line
  corresponds to baryonic
  isodensities for the one-field MOND and the dot-dashed line to the baryonic
  isodensities for the multi-field TeVeS. For the same potential, the
  typical density in TeVeS is slightly higher than in MOND near the centre,
  and slightly lower in the outskirts.}
\protect\label{fig:galp}
\end{figure*}

\subsection{MOND formula approximation}\label{sec:miglformapp}

In the case of the multi-field TeVeS gravity, if one subtracts the true
gravity $g_0$ in the equatorial plane from the one that we would derive
when applying Eq.(1) to the Newton-Einstein gravity, we obtain the
correction to the rotation curve due to the curl field:
\beq\label{eqn:mfa}
\Delta_s={g_{s,r}(0)\left(1+{g_{s,r}(0) \over 1-\alpha g_{s,r}(0)}\right) -g_0 \over g_0}.
\eeq
When this number is negative, the true MOND force is larger than the one
derived using Eq.(1), while the opposite is true when this
number is positive.

In the case of the one-field MOND gravity, the actual Newtonian
gravitational force $g_N(c)$ can be calculated by solving the Newtonian Poisson
equation for the density $\rho_M(r,\theta)$. Then, replacing $g_{s,r}(0)$
by $g_0 - g_N(0)$ in Eq.(\ref{eqn:mfa}) yields $\Delta_M$, the correction to the
rotation curve due to the curl field in the non-relativistic MOND. Although the two values do not correspond to the
same mass density, the relatively large values of the difference imply
non-negligible differences. Those values are listed in Table 1 for three
different values of $g_0$ ($a_0/2$, $a_0$, $2a_0$), corresponding to the
intermediate gravity regime, and for many different interpolating
functions. We thus showed that it is necessary to do rigourous calculations using the scalar field prescription if one wants to
address the intermediate regime quantitatively in complex geometries.
 
\begin{table*}
\caption{Shows the correction to the rotation curve due to the curl field for six $\alpha$'s (cf. Eq. \ref{eqn:mux}) and for three representative gravity strengths ($g_0=a_0/2$, $a_0$
and $2a_0$). $\epsilon=0.8$.}
\begin{tabular}{|c|ccc|ccc|ccc||}
\hline\hline
$g_0$ $\rightarrow$&&$a_0/2$&&&$a_0$&&&$2a_0$&\\
\hline\hline
$\alpha$ $\downarrow$&$\Delta_s$&$\Delta_M$&$\Delta_s - \Delta_M$&$\Delta_s$&$\Delta_M$&$\Delta_s - \Delta_M$&$\Delta_s$&$\Delta_M$&$\Delta_s - \Delta_M$\\
\hline
0.0 &0.889 &-0.782&      1.671&    -0.562&     0.537&     -1.100&    -0.307&    -0.223&   -0.084\\
0.2 &-0.044 &-0.032&   -0.011&     0.525&     -1.130&      1.654&    -0.648&      1.375&     -2.024\\
0.4&3.001 &-1.627&      4.628& -0.197&    -0.123&   -0.073&    -0.807&      5.808&     -6.616\\
0.6 &1.002 &-0.895&1.898&    -0.564&      1.167&    -1.731&    -0.608&    -0.334&    -0.274\\
0.8 &0.002 &0.002&0.000&   0.007 &     -1.238&      1.244&    -0.731&     -7.074&      6.343\\
1.0 &-0.114 &-1.985&1.871&    -0.255&    -0.102&    -0.152&    -0.820&     -2.132&      1.311\\
\hline\hline
\end{tabular}
\end{table*}

\section{Conclusion}\label{sec:conc}
In summary, we have presented analytical models of dynamics and lensing in MOND/TeVeS.
\\
\\
1. We proposed (\S \ref{sec:inter}) a useful set of interpolating functions for MOND, with
a physical counterpart in TeVeS, contrary to the standard interpolating
function commonly used to fit galactic rotation curves.
\\
\\
2. Using our interpolating functions, we found a useful family of spherical
 (\S \ref{sec:spherpot}) models in MOND/TeVeS, with moderate $1/r$
 cusps. Those models were then extended to non-spherical ones, by
 considering multi-centred models (\S \ref{sec:twocenter}), flattened
 models (\S \ref{sec:flatmond})  and scale-free flattened models (\S
 \ref{sec:scalefree}).
\\
\\
3. We showed that the lensing and the orbits in these spherical or
flattened models are rather similar to the expectation in Einstein-Newton
gravity, but still trigger a few surprises in extreme geometries. In
multi-centred models, the convergence map does not always reflect the
projected matter in the lens plane in MOND. This cautions simple interpretations of the analysis of weak lensing in the bullet cluster 1E 0657-56 (Clowe
et al. 2004, see Fig. \ref{fig:bullet}).
\\
\\ 
4. In flattened scale-free models (\S \ref{sec:scalefree}) we also found that there are differences in the
potential-density pairs between the
single-field MOND formulation of Bekenstein \& Milgrom (1984) and the
multi-field TeVeS formulation (Bekenstein 2004). However, the differences are probably not sufficient to solve
the MONDian mass discrepancy in galaxy clusters (Sanders 2003) as suggested
by Bekenstein (2005), unless other parameters of the TeVeS theory, e.g.,
$\Xi$ (see Eq. \ref{eqn:xi}), are important. 
\\
\\
Note that our results are not inconsistent with the
well-known result that rotation curves of astronomical disks are
insensitive to the detailed formulation of MOND. While the very squashed axisymmetric systems of \S \ref{sec:scalefree} are not
very realistic by themselves, they seem to suggest
that the three versions of
MOND are likely to show greater differences in systems of complex
multi-centred geometry, which is realistic for systems undergoing mergers. \footnote{For example, satellites in the tidal field of
a galaxy have interesting Roche lobe shapes, which contain
detailed information on the different laws of gravity (Zhao \& Tian,
2006).} Altogether we would argue
that the most promising systems to test different versions of MOND
are systems of lower symmetry for which MOND was not designed and
the least is known. We have shown here
that application of this test requires highly non-trivial computation to be
done {\it properly}.

\begin{appendix}

\section{Isotropic Distribution Function}

Here we present the full isotropic distribution function corresponding to the
models of \S \ref{sec:spherpot}. We start from the Eddington's formula (see Binney \& Tremaine 1987) 

\beq
F(E)={1 \over \sqrt{8} \pi^2} \left[\int_{E}^{\infty} {d^2\rho \over d{\Phi}^2} {d\Phi \over \sqrt{\Phi - E}}+
{1 \over \sqrt{\Phi- E}}\left({d\rho \over d\Phi}\right)_{\Phi=\infty}\right].
\eeq
Initially we define $\rho$ as a function of $\Phi$
\beq
\rho(\Phi)={M_0 \lambda^2 ((2p+1)\lambda^{-1}+1)  \over 4\pi p^2r_0^3(\lambda^{-1}-1)(p\lambda^{-1}+1)^2}, \qquad \lambda = \exp{-\Phi \over v_0^2}.
\eeq
The first derivative of $\rho$ w.r.t. $\Phi$ is
\beq
{2\pi r_0^3 v_0^2 \over M_0}{d\rho \over d\Phi}={-2p(2p+1)\lambda^{-1} + (3p^2-3p-1)+(3p-1)\lambda+\lambda^2\over (\lambda^{-1}-1)^2(p\lambda^{-1}+1)^3}.
\eeq
Evaluated at $\Phi=\infty$ we get ${d\rho \over d\Phi}|_{\Phi=\infty}=0$.
The second derivative w.r.t. $\Phi$ is
\beq
{2\pi v_0^4r_0^3 \over M_0}{d^2\rho \over d\Phi^2}={8p^2(2p+1)\lambda^{-3}+p(-23p^2+15p+7)\lambda^{-2}
+(9p^3-29p^2+11p+2)\lambda^{-1}+(12p^2-20p+3)+(8p-5)\lambda+2\lambda^{2} \over p^2(\lambda^{-1}-1)^3(p\lambda^{-1}+1)^4}.
\eeq
The reduced Eddington's formula
\beq
F(E)={1 \over \sqrt{8} \pi^2} \left[\int_{E}^{\infty} {d^2\rho \over d{\Phi}^2} {d\Phi \over \sqrt{\Phi - E}}\right].
\eeq
can now easily be numerically integrated as shown in Fig.(\ref{fig:isodf}).
\end{appendix}
\bsp

\label{lastpage}

\end{document}